\RequirePackage[l2tabu,orthodox]{nag} 
\documentclass[12pt,letterpaper]{article} 

\usepackage[T1]{fontenc} 
\usepackage[utf8]{inputenc} 
\usepackage{ebgaramond} 
\usepackage{tgheros} 

\usepackage[USenglish]{babel} 
\usepackage[strict,autostyle]{csquotes} 
\usepackage[babel=true]{microtype} 

\usepackage{amsmath} 
\usepackage{authblk} 
\usepackage{booktabs} 
\usepackage{caption} 
\usepackage{datetime} 
\usepackage[final]{draftwatermark} 
\usepackage{endnotes} 
\usepackage{geometry} 
\usepackage{graphicx} 
\usepackage{natbib} 
\usepackage{rotating} 
\usepackage{setspace} 
\usepackage{titlesec} 
\usepackage{url} 

\usepackage{hyperref} 
\usepackage{orcidlink} 

\newdateformat{monthyeardate}{\monthname[\THEMONTH] \THEYEAR}

\newcommand{\myname}{Geoff Boeing}

\newcommand{\myorcid}{0000-0003-1851-6411}  
\newcommand{\myaffiliation}{Department of Urban Planning and Spatial Analysis\\University of Southern California}
\newcommand{\paperdate}{\monthyeardate\today}
\newcommand{\papertitle}{Urban Science Beyond Samples: Up-to-Date Street Network Models and Indicators for Every Urban Area in the World}
\newcommand{\papercitation}{Boeing, G. 2026. \papertitle. \textit{Environment and Planning B:\@ Urban Analytics and City Science}, published online ahead of print. \href{https://doi.org/10.1177/23998083261446991}{doi:10.1177/23998083261446991}}
\newcommand{\paperkeywords}{Urban Planning, Transportation, Data Science}

\graphicspath{{./figures/}}

\titleformat{\section}{\normalfont\sffamily\large\bfseries\color{black}}{\thesection.}{0.3em}{}
\titleformat{\subsection}{\normalfont\sffamily\small\bfseries\color{black}}{\thesubsection.}{0.3em}{}
\titleformat{\subsubsection}{\normalfont\sffamily\small\color{black}}{\thesubsubsection.}{0.3em}{}

\hypersetup{
    pdfauthor={\myname},
    pdftitle={\papertitle},
    pdfsubject={\papertitle},
    pdfkeywords={\paperkeywords},
    pdffitwindow=true, 
    breaklinks=true, 
    colorlinks=false, 
    pdfborder={0 0 0} 
}

\begin{document}

\title{\papertitle\footnote{{Citation info: \papercitation}}}
\author[]{\myname~\orcidlink{\myorcid}}
\affil[]{\myaffiliation}
\date{\paperdate}

\maketitle

\begin{abstract}

Urban planners need up-to-date, global, and consistent street network models and indicators to measure resilience and performance, model accessibility, and target local quality-of-life interventions. This article presents up-to-date street network models and indicators for every urban area in the world. It uses 2025 urban area boundaries from the Global Human Settlement Layer, allowing users to join these data to hundreds of other urban attributes. Its workflow ingests 180 million OpenStreetMap nodes and 360 million OpenStreetMap edges across 10,351 urban areas in 189 countries. The code, models, and indicators are publicly available for reuse. These resources unlock worldwide urban street network science beyond samples as well as local analyses in under-resourced regions where models and indicators are otherwise less-accessible.

\end{abstract}

\section{Introduction}

Scholars and practitioners use spatial graphs to model street networks to understand or predict urban phenomena including traffic dynamics, accessibility to daily living needs, and the resilience and sustainability of the urban form \citep{barthelemy_spatial_2022}. These networks are defined by both their topology---e.g., connections and configuration---and geometry---e.g., positions, lengths, areas, and angles \citep{barthelemy_spatial_2011}. Accordingly, various topological and geometric indicators exist throughout the literature to measure important street network characteristics. For instance, node degrees reveal streets' connectedness \citep{barrington-leigh_global_2020}, weighted betweenness centralities identify relatively important parts of the network \citep{barthelemy_universal_2025}, and circuity indicates its efficiency or lack thereof \citep{giacomin_road_2015}. Such indicators can then inform downstream urban analytics to target planning interventions or benchmark and monitor cities' progress toward stated sustainability goals \citep{higgs_global_2024}.

Up-to-date, global, consistent urban street network models and indicators are needed now more than ever before as planners face intertwined sustainability and public health crises in cities around the world \citep{giles-corti_creating_2022}. Meanwhile, urban science seeks to expand beyond the limits of traditional sampling to build universal theory and better understand under-studied regions, such as the Global South \citep{lobo_urban_2020}. Yet the limitations of traditional data sources and methods present headwinds to these efforts. Data on urban streets are often digitized inconsistently from place to place, thwarting apples-to-apples global comparisons and making analyses particularly difficult in under-resourced regions \citep{liu_generalized_2022}. Popular data sources such as OpenStreetMap offer reasonably high quality data around the world, but do not package it in graph-theoretic form nor provide indicators \citep{boeing_modeling_2025}. Tools like OSMnx help, but still require coding knowledge to conduct the analysis and potentially extensive computational resources for someone trying to conduct global urban science.

This article presents a resource to fill this gap: street network models and indicators worldwide---plus reproducible code to regenerate them---for scholars and practitioners to easily reuse without reinventing the wheel. Using data from OpenStreetMap and boundaries from the 2025 Global Human Settlement Layer (GHSL), this study models and analyzes the street networks of every urban area in the world, ingesting 180 million OpenStreetMap nodes and 360 million OpenStreetMap edges across 10,351 urban areas in 189 countries. The rest of this article describes this open data repository of street network models and indicators, as well as the open-source software repository containing the code to generate them.

\section{Reproducible Methods}

The following computational workflow, written in the Python programming language, generates the models and calculates the indicators.

\subsection{Urban Boundaries}

The workflow begins by extracting the boundary polygons of each urban area in the world from the 2025 GHSL Urban Centre Database (UCD), which contains 11,422 entities.\ \citet{mari_rivero_urban_2025} describe this input dataset in detail, but to summarize, the GHSL integrates a vast array of census data, remote sensing data, and volunteered geographic information to delineate the world's urbanized areas' boundaries and attach corresponding attribute data. We retain urban areas with >1 km\textsuperscript{2} built-up area and a \enquote{high} GHSL quality control score, resulting in 10,351 urban areas. This basic filtering ensures that we are modeling true urbanized areas rather than false positives or small villages.

\subsection{Network Modeling}

We use OSMnx v2.0.2 to download OpenStreetMap raw data in February 2025 then construct a spatial graph model of the street network within each urban area. The models' theoretical framework is detailed in \citet{boeing_topological_2025}, but we define a \textit{street} as a public drivable thoroughfare in an urban built environment. Thus, this includes cyclist and pedestrian infrastructure that is part of the street network itself, but does not include other paths such as passageways through buildings, footpaths or equestrian trails through nature preserves, or cycleways through parks, as these are not \enquote{streets} by common definition. However, this code workflow provides free, flexible reuse by anyone, including adjusting the filters to easily model different transport networks by re-running the code for any other network type.

The resulting models are nonplanar directed multigraphs with possible self-loops \citep{boeing_modeling_2025}. They have node/edge attribute data from OpenStreetMap, plus geographic coordinates and geometries, and we parameterize OSMnx to retain all graph components and run its edge simplification algorithm \citep{boeing_topological_2025}. Each urban area's graph is saved as a GraphML file, a standard graph serialization format.

\subsection{Elevation}

We attach elevation, in meters above sea level, to each node in each urban area's graph using two global digital elevation models (GDEMs): the Advanced Spaceborne Thermal Emission and Reflection Radiometer (ASTER) v3 GDEM, and the Shuttle Radar Topography Mission (SRTM) version 3.0 GDEM with voids filled. Both are 1 arcsecond (approximately 30-meter) resolution. We download all the GDEM raster files for ASTER (45,824 tiles) and SRTM (14,297 tiles) from NASA EarthData, build a virtual raster for each, then use OSMnx to load each GraphML file and attach the elevation from ASTER and SRTM to each graph node.

We choose one elevation value to use as the \enquote{official} node elevation by comparing the ASTER and SRTM values to a \enquote{tie-breaker} value from Google. To do so, we download each node's elevation from the Google Maps Elevation API then choose whichever value is nearer to Google's. Finally we calculate edge grades and re-save each GraphML file with these node/edge attributes.

\begin{table}[htbp]
    \centering
    \footnotesize
    \caption{The indicators dataset contents. Variables carried over from GHSL are noted.}\label{tab:indicators}
    \begin{tabular}{p{3.8cm} p{1.2cm} p{7.6cm}}
        \toprule
        Variable                      & Type    & Description \\
        \midrule
        area\_km2 & integer & Area within urban center boundary polygon, km2 (GHSL) \\
        bc\_gini & decimal & Gini coefficient of normalized distance-weighted node betweenness centralities \\
        bc\_max & decimal & Max normalized distance-weighted node betweenness centrality \\
        built\_up\_area\_m2 & integer & Built-up surface area, square meters (GHSL) \\
        cc\_avg\_dir & decimal & Average clustering coefficient (unweighted/directed) \\
        cc\_avg\_undir & decimal & Average clustering coefficient (unweighted/undirected) \\
        cc\_wt\_avg\_dir & decimal & Average clustering coefficient (weighted/directed) \\
        cc\_wt\_avg\_undir & decimal & Average clustering coefficient (weighted/undirected) \\
        circuity & decimal & Ratio of street lengths to straightline distances \\
        core\_city & string & Urban center core city name \\
        country & string & Primary country name \\
        country\_iso & string & Primary country ISO 3166--1 alpha--3 code \\
        elev\_iqr & decimal & Interquartile range of node elevations, meters \\
        elev\_mean & decimal & Mean node elevation, meters \\
        elev\_median & decimal & Median node elevation, meters \\
        elev\_range & decimal & Range of node elevations, meters \\
        elev\_std & decimal & Standard deviation of node elevations, meters \\
        grade\_mean & decimal & Mean absolute street grade (incline) \\
        grade\_median & decimal & Median absolute street grade (incline) \\
        intersect\_count & integer & Count of (undirected) edge intersections \\
        intersect\_count\_clean & integer & Count of street intersections (merged within 10 meters geometrically) \\
        intersect\_count\_clean\_topo & integer & Count of street intersections (merged within 10 meters topologically) \\
        k\_avg & decimal & Average node degree (undirected) \\
        length\_mean & decimal & Mean street segment length (undirected edges), meters \\
        length\_median & decimal & Median street segment length (undirected edges), meters \\
        length\_total & decimal & Total street length (undirected edges), meters \\
        node\_count & integer & Count of nodes \\
        orientation\_entropy & decimal & Entropy of street network bearings \\
        pagerank\_max & decimal & The maximum PageRank value of any node \\
        prop\_4way & decimal & Proportion of nodes that represent 4-way street intersections \\
        prop\_3way & decimal & Proportion of nodes that represent 3-way street intersections \\
        prop\_deadend & decimal & Proportion of nodes that represent dead-ends \\
        resident\_pop & integer & Total resident population (GHSL) \\
        self\_loop\_proportion & decimal & Proportion of edges that are self-loops \\
        straightness & decimal & Inverse of circuity \\
        street\_segment\_count & integer & Count of streets (undirected edges) \\
        uc\_id & integer & Urban center unique ID (GHSL) \\
        uc\_names & string & List of city names within this urban center (GHSL) \\
        world\_region & string & UN SDG geographic region \\
        \bottomrule
    \end{tabular}
\end{table}

\subsection{Indicator Calculation}

For each graph, we calculate the street network indicators described in Table~\ref{tab:indicators}. These include geometric and topological measures common in transport planning, urban design, and network science \citep{barthelemy_spatial_2011,barthelemy_spatial_2022,boeing_street_2022}. We report node counts, intersection counts (i.e., non-dead-end nodes), and both geometrically and topologically consolidated intersection counts, as explained in \citet{boeing_topological_2025}. We also calculate betweenness centrality for every node in every graph. A node's betweenness centrality measures the share of all possible shortest paths in a graph that use that node. High centrality values indicate \enquote{important} nodes relied on by many shortest paths \citep{boeing_urban_2024}. The maximum betweenness centrality represents the highest relative value in a graph (and thus identifies the most important node), and their Gini coefficient measures the concentration of \enquote{importance} in a network, indicating the presence and severity of chokepoints \citep{barthelemy_universal_2025}.

\subsection{Data Repository Preparation}

We convert each GraphML file to a GeoPackage and node/edge list files. The former allows users to work with these spatial networks in any GIS software. The latter provides a minimal, lightweight, highly compressible version of the models. Then we perform a series of file verification checks and create metadata files for the graphs' node and edge attributes and all of the indicators. Finally we compress and upload all model files (GeoPackages, GraphML, and node/edge lists), indicators, and metadata to the Harvard Dataverse.

\section{Code and Data Products}

\subsection{Code Repository}

The preceding methods are fully reproducible by running the modeling and analytics workflow made publicly available in the source code repository\endnote{Code repository: https://github.com/gboeing/street-network-models} on GitHub under the terms of the MIT license. A well-equipped personal computer can execute this workflow, but given the resource requirements it may be better (and faster) to run it in a high-performance computing cluster, where available. The input data, dependencies, and resources required to run it are documented in the code repository.

\subsection{Data Repository}

The data repository comprises five datasets nested within a top-level Dataverse\endnote{Top-level Dataverse: https://dataverse.harvard.edu/dataverse/global-urban-street-networks} repository:

\begin{itemize}
    \item Global Urban Street Networks GeoPackages\endnote{Global Urban Street Networks GeoPackages: https://doi.org/10.7910/DVN/E5TPDQ}
    \item Global Urban Street Networks GraphML files\endnote{Global Urban Street Networks GraphML files: https://doi.org/10.7910/DVN/KA5HJ3}
    \item Global Urban Street Networks Node/Edge lists\endnote{Global Urban Street Networks Node/Edge lists: https://doi.org/10.7910/DVN/DC7U0A}
    \item Global Urban Street Networks Indicators \endnote{Global Urban Street Networks Indicators: https://doi.org/10.7910/DVN/ZTFPTB}
    \item Global Urban Street Networks Metadata \endnote{Global Urban Street Networks Metadata: https://doi.org/10.7910/DVN/WMPPF9}
\end{itemize}

The model files are zipped at the country level, and each is identified by its urban area name and UCD ID.\ The latter allows users to join these indicators to GHSL attribute data.

\section{Practical Applications}

In terms of downstream applications, these models and indicators can be used to estimate relationships between network topology and travel patterns or land use \citep{wang_road_2020,wang_street_2018}, identify relatively important parts of a network via empirical distributions of within-city betweenness centrality \citep{barthelemy_universal_2025}, measure sprawl \citep{gervasoni_calculating_2017}, simulate disasters' effects on street network resilience around the world \citep{boeing_resilient_2024}, improve logistics' last-mile trip predictions \citep{merchan_quantifying_2020}, and estimate relationships between greenhouse gas emissions and street network design \citep{boeing_urban_2024}. To date, this repository's models and indicator data have been downloaded over 70,000 times by users.

\citet{dyer_commentary_2024} argue that urban scholars and practitioners need models and indicators that keep up with the pace of transformational change in an era of rapid urbanization. This project seeks to address this by building on prior work initially conducted in 2019--2020 that generated a preliminary version of the data repository \citep{boeing_street_2022}. That initial version was based on the 2015 version of the GHSL UCD and 2020 OpenStreetMap data. This new version uses the 2025 GHSL UCD and 2025 OpenStreetMap data to make several new contributions.

First, it includes over 1,400 more urban areas and 11 more countries than the earlier version. This entails significantly more worldwide coverage in an era of rapid urban expansion. That is, these new models allow us to study more---and more-recently urbanized---cities than before.

Second, these new models incorporate 10 years of recent urbanization into their updated urban area boundaries and 5 years of new community additions to OpenStreetMap. As such, this workflow's modeling includes approximately 20 million more street network nodes and 40 million more edges than the earlier version. The new urban boundaries allow users to link these street network models and indicators to hundreds of new, up-to-date GHSL attributes on urban climate, land use, economic conditions, etc. For instance, this allows us to simulate urban travel and access in the context of extreme heat or cold, generate integrated models of urban networks and land uses, or estimate relationships between economic activity and underlying transport infrastructure.

Third, it adds new attributes and indicators to the repository---most consequentially the betweenness centrality of every node in every urban area's street network, which is time- and resource-intensive to calculate, yet unlocks useful analyses of network structure and resilience. For example, betweenness centrality can provide us with indicators of chokepoints when simulating network disruptions and resilience \citep{boeing_resilient_2024} and indicators of the relative importance of network elements \citep{barthelemy_universal_2025}. Fourth, it uses finer-grained SRTM data (30m instead of the previous 90m resolution) for more precise elevation attribute values. This is especially useful when modeling active travel behavior, which is sensitive to arduous elevation changes \citep{barthelemy_surfacic_2024}.

Fifth, from a \textit{code product} perspective, the workflow's code base has been wholly refactored and rewritten from the ground-up to significantly reduce its cyclomatic complexity, memory use, and runtime. This makes the workflow more maintainable, sustainable, and easier to re-run in the future to periodically update the data repository whenever new GHSL data are released. This also makes it easier for downstream users to re-run the code to model any network types they wish---including pedestrian networks or cycling networks, for example.

Finally, these models and indicators themselves unlock other researchers' work, such as through the example applications discussed above. This project provides a global dataset to conduct both worldwide urban street network science beyond samples as well as local analyses particularly in less-resourced regions where such models and indicators are most needed, yet most scarce \citep{giles-corti_creating_2022}.

\section{Data Availability}

Users may freely download the models or indicators directly from the aforementioned Dataverse, or access the source code and documentation at the aforementioned GitHub source code repository.

\IfFileExists{\jobname.ent}{\theendnotes}{}

\setlength{\bibsep}{0.00cm plus 0.05cm} 
\bibliographystyle{apalike}
\bibliography{references}

\end{document}